\documentclass[journal]{IEEEtran}
\usepackage{cite}
\usepackage{balance}

\usepackage{color}
	\usepackage{soul}
\ifCLASSINFOpdf
   \usepackage[pdftex]{graphicx}
\else
   \usepackage[dvips]{graphicx}
\usepackage[normalem]{ulem}
   
\fi
\usepackage{multirow}
\usepackage{rotating} 

\begin{document}
%
\title{Nature-Inspired Computational Model of Population Desegregation under Group Leaders Influence}
%
%
%

\author{Kashif~Zia,~
        Dinesh~Kumar~Saini,~
        Arshad~Muhammad,~
        and Alois~Ferscha~\IEEEmembership{Member,~IEEE}
\thanks{K. Zia is with the Faculty of Computing and Information Technology, Sohar University, Oman e-mail: kzia@soharuni.edu.om.}
\thanks{D.K. Saini, and A. Muhammad are with the Faculty of Computing and Information Technology, Sohar University, Oman.}
\thanks{Alois Ferscha is with the Institute of Pervasive Computing, Johannes Kepler University, Linz, Austria.}
\thanks{Manuscript received April 19, 2005; revised August 26, 2015.}}

%
%

\markboth{Journal of \LaTeX\ Class Files,~Vol.~14, No.~8, August~2017}%
{Zia \MakeLowercase{\textit{et al.}}: Population Desegregation and Group Influence}
%



\maketitle

\begin{abstract}
This paper presents an agent-based model of population desegregation and provides a thorough analysis of the social behavior leading to it, namely the contact hypothesis. Based on the parameters of {\it frequency} and {\it intensity} of influence of group leaders on the population, the proposed model is constituted by two layers: (i) a {\it physical layer} of the population that is influenced by (ii) a {\it virtual layer} of group leaders. The model of negotiation and survival of group leaders is governed by the nature-inspired evolutionary process of queen ants, also known as Foundress Dilemma. The motivation of using a virtual grouping concept (instead of taking a subset of population as the group leaders) is to stay focused on finding the conditions leading individuals in a society tolerating a significantly diversified (desegregated) neighborhood, rather than, indulging into complex details, which would be more relevant to studies targeting the evolution of societal group and leaders. A GIS-driven simulation is performed, which reveals that (i) desegregation is directly proportional to the frequency of group leaders' contact with the population, and (ii) mostly, it remains ineffective with an increase in the intensity of group leaders' contact with the population. The mechanism of group selection (the conflict resolution model resolving the Foundress Dilemma) reveals an exciting result concerning negative influence of cooperative group leaders. Most of the time, desegregation decreases with increase in cooperative leaders (the leaders enforcing desegregation) when compared with fierce leaders (the leaders enforcing segregation).      
\end{abstract}


%
\IEEEpeerreviewmaketitle
\vspace{-3mm}
\section{Introduction}

According to Oxford dictionary, segregation is defined as ``the action or state of setting someone or something apart from others''. Historically, segregation is connected with the dynamics of biological, and more specifically human population. For a human population, segregation is a progressive outcome of population demography as a result of discriminatory settlement behavior of the individuals \cite{schelling1971dynamic}. The discrimination is based on some bias, such as sex, race or color that influences people decision of where to live. It has been observed that such individualistic decisions have shaped the demographics of cities, sometimes resulting in exceptionally segregated settlements \cite{white1986segregation}. 
Settlements having extreme segregation are sometimes not desirable. For example, the countries that have experienced racial abuse and ethnic wars desire desegregation (anti-segregation) in its population\cite{gona2012dawn}. Brewer \cite{brewer2013social} defined desegregation as: ``extended contact between previously isolated social groups that are brought on either by acute or gradual processes of change''. However, this change cannot be brought about through legislation and laws \cite{christopher2005slow}. It is established from the experience that such a change is inseparable without people themselves realize change \cite{oldfield2004urban} \cite{liu2005school}. Human societies evolve with time may be due to the intention of people to mix-up and desegregate. Standard processes of evolution are randomness, agency, organization, and contingency \cite{Evolution2017}. Agent-based Modeling (ABM) provides an ideal platform to model and configure these processes \cite{macy2002factors}, thus acting as an enabler not only to observe the evolutionary process itself but also, find answers of the interesting ``what-if'' questions.

A model, whether agent-based or not is always based on a foundation. For social phenomena, such as segregation (desegregation), this can be based on a theory or empirical evidence. Since empirical evidence of individualistic behavior in the population is difficult to get and synthesize towards a population-level dynamics, the theory-driven foundations are often more appropriate. However, theories are intrinsically contradictory. Computer simulation provides means to evaluate the contradictory theories relating to the same phenomena and specify conditions in which one theory is applicable, while others are not.

The rapidly growing field of Computational Social Science (CSS) \cite{gilbert2010computational} \cite{wallach2016computational} is about understanding a society using the computer simulation. More recent tools, such as Cellular Automata (CA) (from physics and mathematics), and Distributed Artificial Intelligence and agent technology (from computer science) have influenced this growth positively \cite{conte2012manifesto}. Due to its nature, the social simulation should be a theory-driven system with a focus on the explanation of the phenomena of interest (rather than the desire for specific outcomes) \cite{cioffi2017computation}. The novelty of social simulation is based on the observation that an overwhelmingly complex behavior emerges from relatively simple local activities \cite{simon1996sciences}. ABM is a popular modeling technique used by CSS researchers due to its ability to reproduce the societal effects purely based on interactions at the local level \cite{gilbert2005simulation}. CSS has a long history and a strong underpinning on the social theories.

Schelling's seminal computational model of the population segregation \cite{schelling1971dynamic} has been a reference point for diverse research streams \cite{easley2010networks}, such as, population dynamics and land use \cite{hwang2014divergent, bayer2016dynamic, parker2003multi, massey1993american, henderson2014economic}, social interaction dynamics \cite{castellano2009statistical}, social and psychological behavior studies\cite{akerlof1980theory} \cite{crane1991epidemic}, and business and market analysis \cite{easley2010networks}. Shelling's model suggests that even if {\it the majority of people are not biased and tolerate a neighborhood (locality), which is not exactly of their liking, segregation is still a possibility on a global scale}. More specifically, Schelling's model is successful in demonstrating that a predominantly tolerant population may also generate a segregated settlement, thus, validating that what emerges at a global level may not be a true representation of the local parts. The strength of otherwise a basic neighborhood-based model was vested in agents' strict localized decisions producing an unexpected outcome at the population level. Schelling's model, therefore, was one of the first experimental evidence for the emergence of totally unpredictable global behavior, which was contradictory to the local rules producing it. Due to this reason, Schelling's model is often used to demonstrate the strengths of an agent-based modeling paradigm; a modeling technique specifically developed to analyze these situations.

To model a socially-inspired model of desegregation, it is important to understand the human behavior and the social processes involved.  Norman S. Miller and Marilynn B. Brewer defined a relevant hypothesis, that is the {\bf contact hypothesis} as: ``the idea that prejudice and hostility between members of segregated groups can be reduced by promoting the frequency and intensity of intergroup contact'' \cite{miller2013groups}. However, there are other theories, which entirely contradict this theory. For example, it is also claimed in \cite{denis2015contact} that ``contact tends to reproduce, rather than challenge, the inequitable racial structure.'' With the help of the model presented in this paper, the conditions in which the contact hypothesis is applicable are evidenced. Consequently, the conditions in which the contact hypothesis is not applicable are also evidenced, thus, favoring the contradictory hypothesis \cite{denis2015contact}.

However, the modality/medium of intergroup contact is hard to resolve. There are numerous factors that can be considered, such as, group boundaries, the scope of influence, relative activism within a group, and the cognitive traits that would be responsible for bringing the change. This could lead to a complex mechanism of defining and evaluating the outcome of intergroup contact. However, in fact, the outcome of intergroup contact influences an individual in a population towards two possible states of minds; appreciation for the societal good (cooperation) or not (defection); provides the much-needed ground for simplification. Therefore, for the purpose of the model presented in this paper, the responsibility of intergroup contact is taken away from the real population and assigned to the artificial society of leaders. It is motivated by the argument that contact itself (without any external anti-segregation impact) is not enough to change the racial structure as suggested in \cite{denis2015contact}. This means that leaders' influence the population to segregate or desegregate based on their type. In this way, the notions of intensity and frequency of intergroup contact are equally replicated as intensity and frequency of leaders' influence on the population.

Hence, the proposed model is based on two layers: a physical layer of real population of {\it agents}, which is influenced by a virtual layer of group {\it leaders}. The periodicity and intensity of the influence exerted by group leaders onto the real population transform it into a settlement, which is segregated or otherwise. Since the real population only decides to segregate or not to segregate, a simple model motivated by Schelling's computational {\bf model of segregation} is sufficient \cite{schelling1971dynamic}. If under this model, an agent does not make a move, even though a majority of its neighbors are not of its kind, it practically acts in favor of desegregation. The model of virtual layer's influence is then the model of interest and termed as the {\bf model of leaders influence}.   

The model of leaders influence on the agents is partially based on the contact hypothesis \cite{miller2013groups}. How much these leaders influence the behavior of agents depends on the attributes related to the leaders, such as (i) capabilities, (ii) relationship, (iii) environment, and (iv) strategies of conflict-resolution (group dynamics). The capabilities are defined by {\it frequency} and {\it intensity} of interaction \cite{miller2013groups} of the leaders with the agents. The relationship between leaders and agents is realized by the presence and proximity constraints, which are in turn dependent on the density of leaders population (an agent must have one or more leaders in its proximity to get itself influenced).
It is notable that the relationship constraints of the environment restrain the capabilities of the leaders. For this paper, the environment is spatial in nature. A CA based environment is used to define the interlayer relationship thus mechanizing the capabilities of interaction between leaders and agents. For other types of environments, this relationship would have been different. For example, if leaders are operating through social networks or mass media, their interaction mode and influence would have been connectivity and availability rather than proximity and presence. 

Lastly, the strategies of grouping and conflict resolution between the leaders are also modeled. A nature-inspired model of grouping, survival, and influence is used. Nature-inspired models provide basic mechanisms of inter-organism interaction leading to mutual survival, role assignment and cooperative problem-solving. For the model proposed in this paper, it makes modeling of leaders straight-forward in terms of: (i) interaction pattern and influence, (ii) relationship and dependability, and (iii) strategies of conflict-resolution. The process of negotiation and survival of the leaders is governed by the nature-inspired evolutionary model of queen ants behavior termed as Foundress Dilemma. 

To the best of our knowledge, the model of such complexity and coverage is not reported previously. The rest of the paper is organized as follows. In section \ref {sec:rw}, a review of related work is given. Section \ref{sec:over} presents an overview of the models. In section \ref {sec:abm}, a detailed description of agent-based models is given. Section \ref {sec:sim} explains the simulation experiments and the analysis of the simulation results, followed by conclusions of the paper in section \ref {sec:conc}.

\vspace{-3mm}
\section {Related Work} \label {sec:rw}

\begin{figure*} 
\centering
\includegraphics[width=0.80\textwidth]{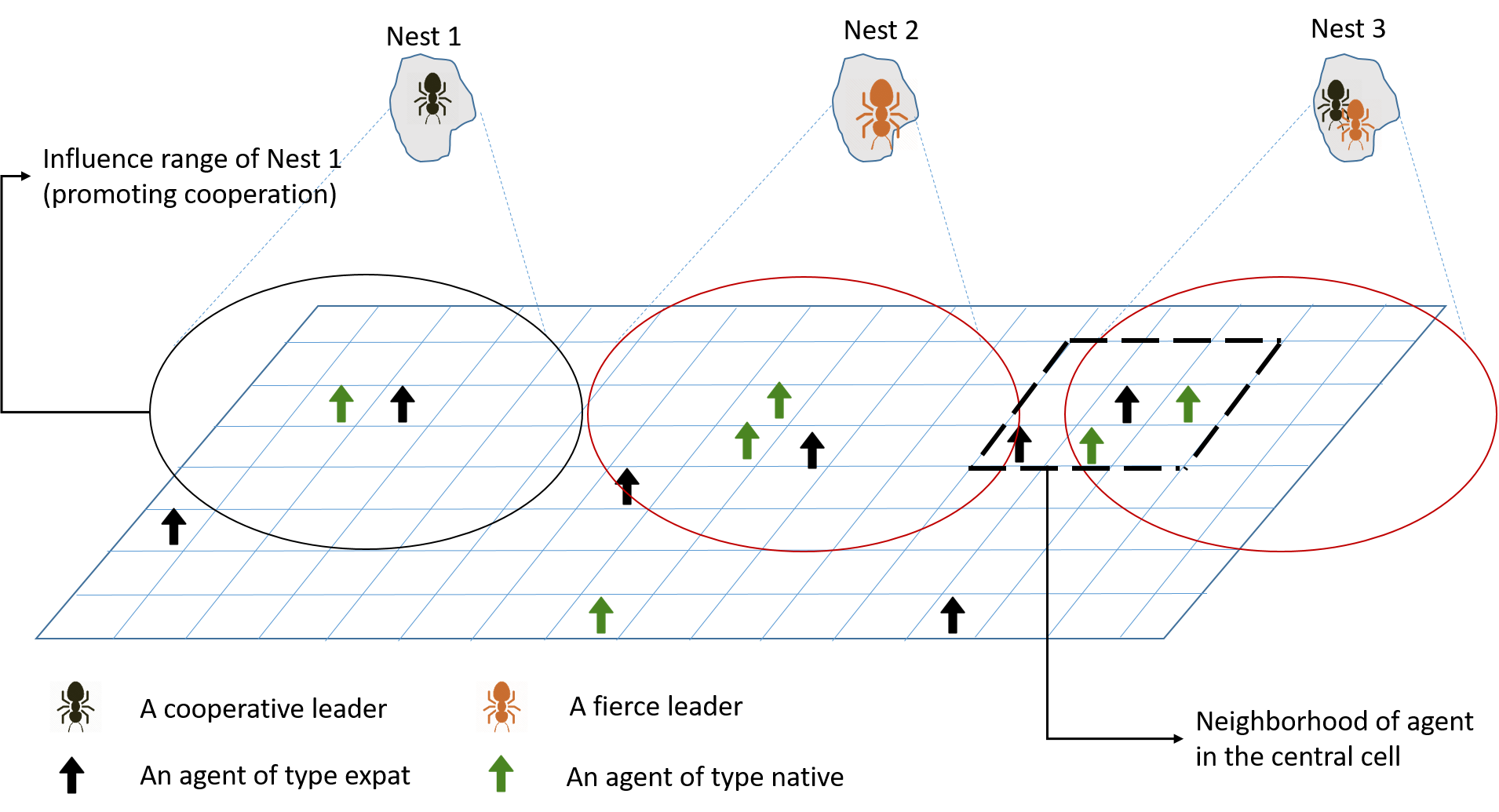}
\caption{A schematic representation of integrated model. The process of segregation works under the combined effect of the neighborhood of an agent and the influence exerted by the leaders. Formation of nests of leaders is governed by a model motivated from Foundress dilemma \cite{shaffer2016foundress}. Each generation of leaders is placed on the map independent of previous placements (this guarantees influence distribution all across the agent population). The frequency of leaders' influence relates to how often they reproduce to produce new generations, whereas, the intensity of leaders' influence relates to how fierce they are (e.g. leader in Nest 2 is more fierce than other leaders, indicated by increased size).}
\label{fig:schematic}
\end{figure*}

Nature-inspired computing is a computing paradigm inspired by self-organizational behavior in naturally-appearing complex systems \cite{liu2006toward}. Originating from the contemporary ant colony \cite{dorigo1992optimization}, and particle swarm \cite{kennedy2011particle} optimization algorithms, more unusual behaviors have been studied more recently \cite{fister2013brief}. These include firefly \cite{yang2010nature}, bat \cite{yang2010new}, and cuckoo search \cite{yang2009cuckoo} algorithms. 

Nigel Gilbert and Klaus G. Troitzsch in their book proposed various social science computational models and simulation methods, which are helpful to understand the society \cite {gilbert2005simulation}. Axelrod is one of the pioneers, who used an evolutionary approach to describe societal norms focusing on the stability and emergence \cite{axelrod1986evolutionary}. A more general discussion on evolutionary algorithms to understand the social processes as a whole is given by Chattoe-Brown, et al. \cite{chattoe1998just}. Similarly, Kenneth A. De Jong presented various dimensions of evolutionary computation in his book, which helps to understand the nature- and bio-inspired models of population behavior \cite{jong2009evolutionary}. Cioffi, De Jong and Bassett in their paper described biologically-inspired approaches to understand complex social systems \cite{cioffi2012evolutionary}. Their work is seminal, because it presents a combined approach of evolutionary approaches used with the agent-based modeling.

An overview of altruism, cooperation, mutualism, strong reciprocity, and group selection in the social settings is provided by West, et al.\cite{west2007social}. A framework for achieving group level goals with the formation of agents' coalitions is presented by Axtell\cite{axtell2002non}. Modeling and simulation work on segregation is widely available \cite{clark1991residential, grauwin2012dynamic, cortez2015attractors}. Romans Pancs, and Nicolaas J Vriend \cite{pancs2007schelling} have adapted Schelling's segregation model to have a strict preference in favor of integration (desegregation). They found that even in such case, the best response dynamics leads to the segregation at a global level. Similar results are also presented by Junfu Zhang \cite{zhang2011tipping}. These are interesting results highlighting that population on its own cannot settle for desegregation; it requires external factors.

Mostly, work on the population desegregation is primarily of empirical nature. For example, Smets, et al. \cite{smets2008together} demonstrated the difficulty in quantifying the population behavior in varying degree of possible contacts. Sahasranaman studied evolutionary dynamics of neighborhood economic status in the cities, which also have a great impact on the desegregation of people in big cities [8]. Hatna, Erez and Besnion studied the Schelling model of ethnic residential dynamics, and their findings suggest an integrated-segregated dichotomy of the patterns in population behavior. Miles Hewstone and Alberto Voci \cite{hewstone2009diversity} have studied the effectiveness of intergroup contact in reducing prejudice in a society. Contrarily, Jeffrey S Denis \cite{denis2015contact} concluded that ``contact tends to reproduce, rather than challenge, the inequitable racial structure''. It is contrary to the expected outcome (desegregation), when applying the contact theory \cite{miller2013groups}. Nevertheless, this is again an empirical study without involving any computational model. 

In this paper, the external driving factor for desegregation is an overlay virtual population representing the influence factors from the leaders. The grouping dynamics of these leaders is mediated by nature-inspired ant queen behavior \cite{shaffer2016foundress}. Queen ants perform nests management following evolutionary mechanism of group selection, which is essential for the survival of the entire population. Hence, using such mechanism to analyze population dynamics (including segregation / desegregation) is natural. Despite its relevance, to the best of our knowledge, no such work exists in the literature. This work is an effort to address this gap.

\section{Model Overview} \label {sec:over}

The model of population segregation is simple, motivated by Shelling's original publication \cite{schelling1971dynamic}. Agents representing human population are of two ethnic {\it type}s; {\it expat} and {\it native}. An agent has a desire to segregate if the majority of the agents in its neighborhood are not of its {\it type}. This contemporary model of segregation is extended to include the influence of leaders in the decision-making of the agents. Hence, an agent would only be happy, if it is happy with its surrounding as well as its compatibility with the type of influence exerted by the leaders in its surrounding. 
The model of leaders influence is motivated from \cite{shaffer2016foundress} and assumes the presence of cognitive leaders (leaders influencing the opinion of the real population to segregate or otherwise). These leaders are virtual because they are not part of the real population. The leaders' behavior is motivated by ant queens' behavior reported in \cite{shaffer2016foundress}, briefly described in the following. 
Ants live in colonies. When young queens reach the age of reproduction, they leave their nest, and try to build their own nest; it is a very challenging task for a queen. In many species of the ants, the queens are predominately {\it fierce}. They fight with other queens and independently build their colony. However, in some species of the ants, the queens, also of being fierce, are also {\it cooperative} and may live in a nest together. In this situation, one of them becomes a leader of the group and performs the duty of reproduction; while others sacrifice their natural instinct of reproduction in favor of the leader queen. In fact, it increases their chances of survival collectively, while adhering to the evolutionary phenomena of group selection. While involved in the process of group selection, the choice to fight or cooperate is defined as Foundress Dilemma \cite{shaffer2016foundress}. 
Motivated from above, there are two types of leaders in the proposed model. {\it Cooperative} leaders are those, who promote a sense of cooperation (possibly leading to desegregation). {\it Fierce} leaders are those who promote a sense of non-cooperation (possibly leading to segregation). The Fierce leaders always fight with the cooperative leaders. The fierce leaders may fight with each other if one or both of them are fierce beyond a threshold. 
The model of leaders influence is comprised of five sequential steps; named as reproduction, clustering, fighting, group competition, and influencing (the details are given in section \ref{sec:abm}). The agent-based model combines the grouping behavior, the dynamics of interaction of the group leaders and the real population, and neighborhood based segregation mechanism; with an underpinning on the environmental constraints. A schematic representation of the overall model is given in Fig. \ref{fig:schematic}. As the population under the influence of a nest (or a leader) is mere of physical nature (agents in influence range of a leader), the outcome (the desegregation index) is entirely depended on the presence, evolution, frequency, and intensity of the influence. This helps to understand the conditions leading to the population desegregation in terms of (i) spread of (de)segregation effort (relative number of cooperative vs. fierce leaders), (ii) evolving leaders' populations and distribution, (iii) intensity of (de)segregation effort (level of fierceness of fierce leaders), and (iv) the frequency of (de)segregation effort (speed of reproduction in the evolutionary process).

\section {Agent-Based Model} \label {sec:abm}

\begin{figure} 
\centering
\includegraphics[width=0.49\textwidth]{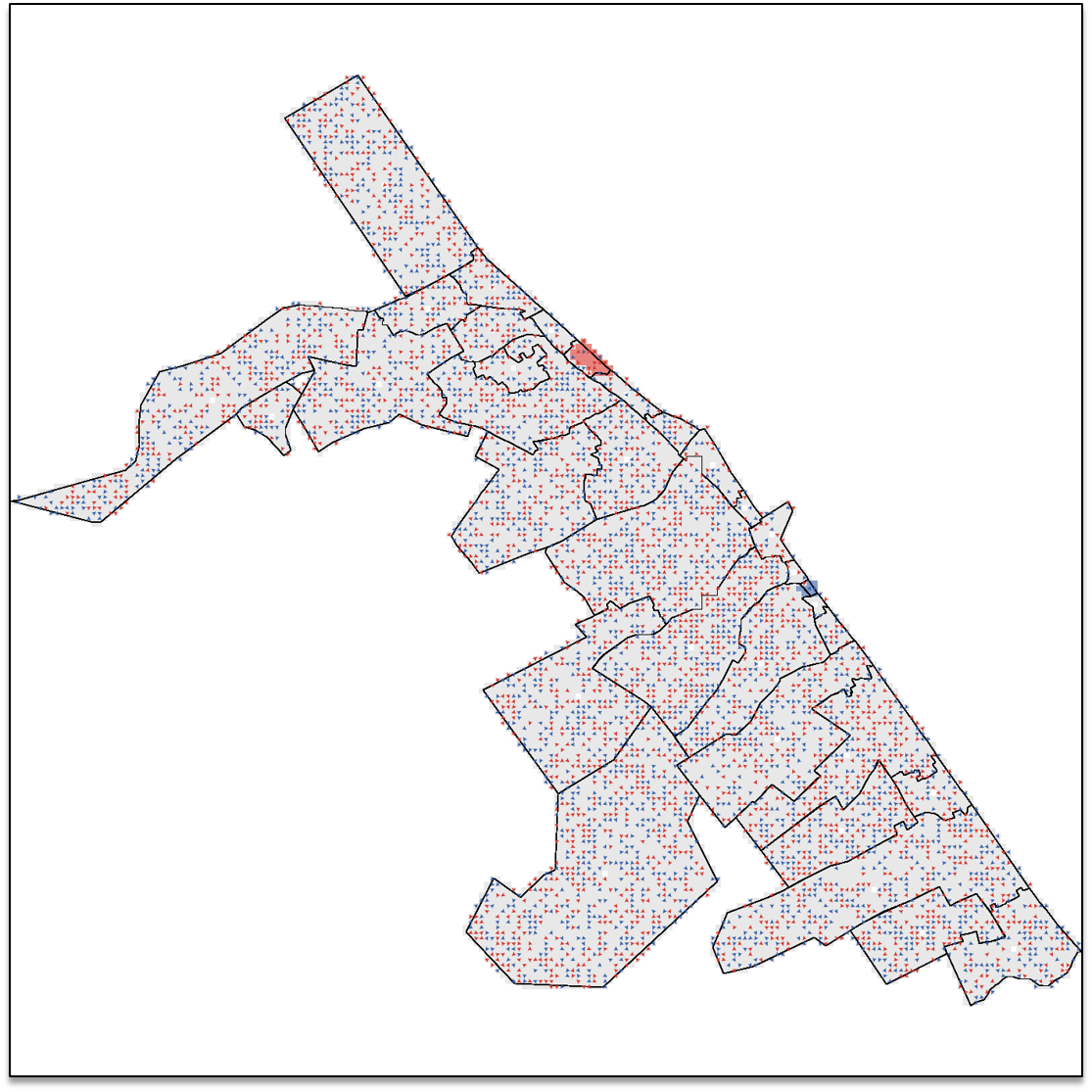}
\caption{Setting up a population of 5000 agents on the map of Sohar city. Agents in blue are expats and agents in red are natives. Two regions, one colored in red and the other colored in blue show the population of these regions segregated enough (segregation is more than segregation threshold of 40\%) in favor of natives and expats, respectively.}
\label{fig:setup}
\end{figure}

The agent-based model proposed in this paper models grouping dynamics, and interaction of the group leaders and the real population with an underpinning on the environmental constraints. Under these settings, the capabilities of group leaders (in terms of frequency and intensity of interaction) towards the agent's decision to stay or leave a place are evaluated. 

\subsection {Simulation Space}
Both, the model of segregation and the model of leaders influence are simulated on the real map. The GIS maps of the city are converted into a CA-based structure. Each cell in the CA {\it World} belongs to a region/locality of the city. Each region of the city has a {\it population}, which is assigned an initial value equal to a fraction of (dependent on the area of the region) total agent population. Each member of this population represents a household and exists as an agent occupying a cell in the corresponding region. The {\it type} of each generated agent is set to either ``expat'' or ``native'', randomly. A region's {\it regionType} can also be one of ``expat'' or ``native'', if region's agent population is segregated enough (defined by prescribed {\it segregation\_threshold}), otherwise it would be ``neutral''. A setup representing these parameters is shown in Fig. \ref{fig:setup}.
\subsection {Model of Segregation} 

\begin{figure} 
\centering
\includegraphics[width=0.49\textwidth]{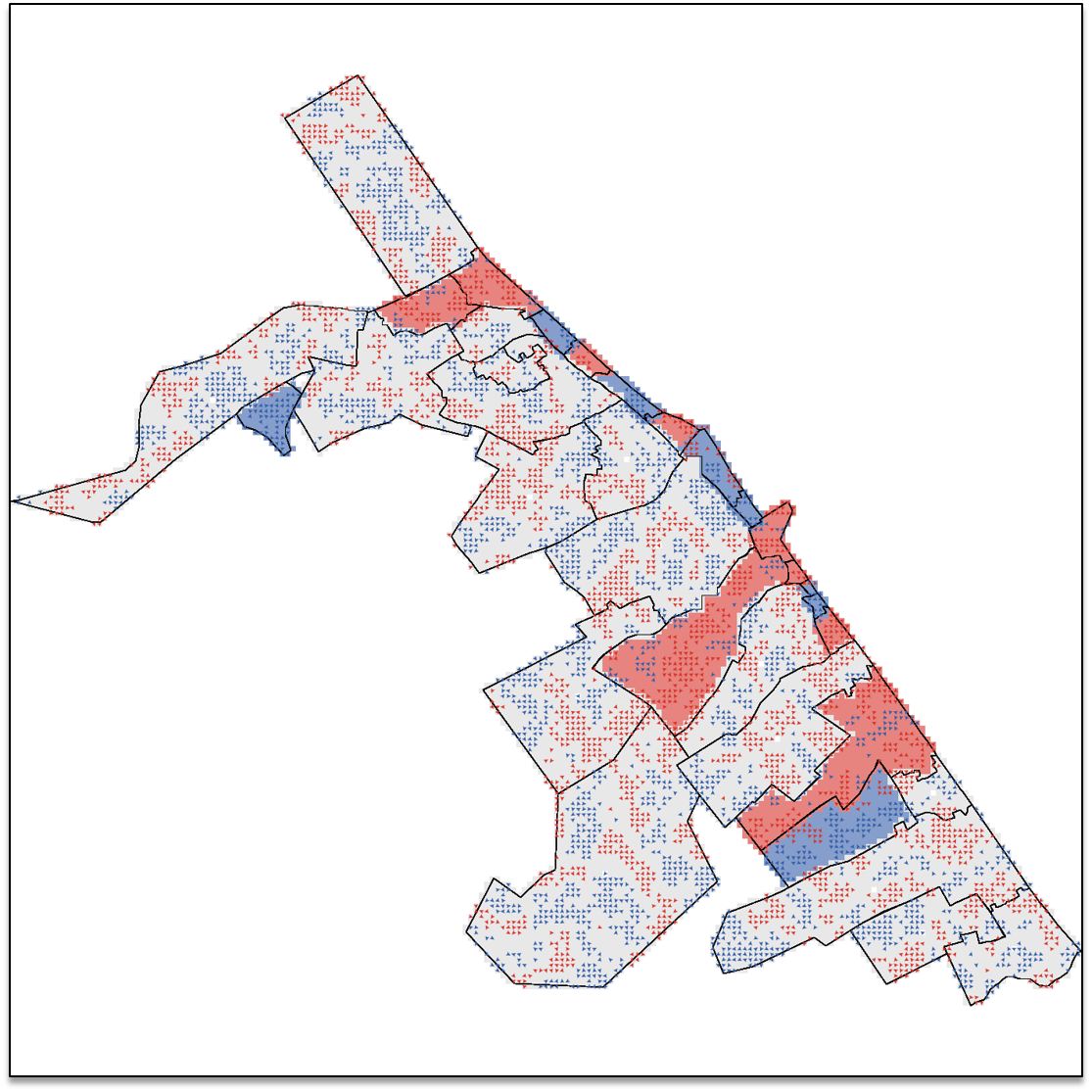}
\caption{Population segregation after equilibrium is achieved at iteration 25. Some regions have attained segregation threshold (region as a whole having regionType ``expat'' [in blue] or ``native'' [in red]) while many regions are still below segregation\_threshold (with Type ``neutral''). Even regions of regionType ``neutral'' achieve local segregation.}
\label{fig:segregate25}
\end{figure}

An agent has a desire to segregate if the majority of agents in its neighborhood are not of its type. The definition of majority is based on the parameter {\it PDTU} (abbreviation for percentage\_difference\_to\_be\_unhappy). If an agent has the ratio of agents in its neighborhood not of its own {\it type} greater than {\it PDTU}, it becomes ``unhappy'' and try to move somewhere else. Otherwise, it is ``happy'' and remains where it is. An unhappy agent will move to a region whose {\it regionType} is equal to its own {\it type} or ``neutral''. Fig. \ref {fig:segregate25} shows population after segregation model is applied with the parameters: agent population = 5000, {\it segregation\_threshold} = 60\%, and {\it PDTU} = 40\%. It is worth noting that the extension of segregation model so that it also incorporates the leaders' influence is given later in the sub-section ``integrated model''.

\subsection {Model of leaders Influence}

A small fraction of cooperative leaders are initially dispersed randomly across the map. Remaining are fierce leaders. These leaders are continuously created, and destroyed in each generation, hence evolving with time. Consequently, the model of segregation gets augmented to incorporate the influence induced by this model. The model of leaders influence operates in five sequential steps, described in the following.\\

\noindent {\bf Reproduce}

At the start of each five-step execution cycle, all the cooperative, and fierce leaders reproduce themselves depending on the previous generation. The population of leaders in each generation is same, but the relation between the cooperative and fierce leaders may change due to mutation. A minimal {\it mutation\_index} equal to $0.01$ would guarantee a change in the population distribution of leaders during evolution. 

\noindent {\bf Cluster}

After reproduction, the leaders if close enough (identified by {\it cluster\_radius}) would cluster, i.e. moving right adjacent (overlapped) to each other irrespective of their type. Hence, several nests would be formed, where each nest may contain more than one leaders of different types.

\begin{figure*} 
\centering
\includegraphics[width=0.75\textwidth]{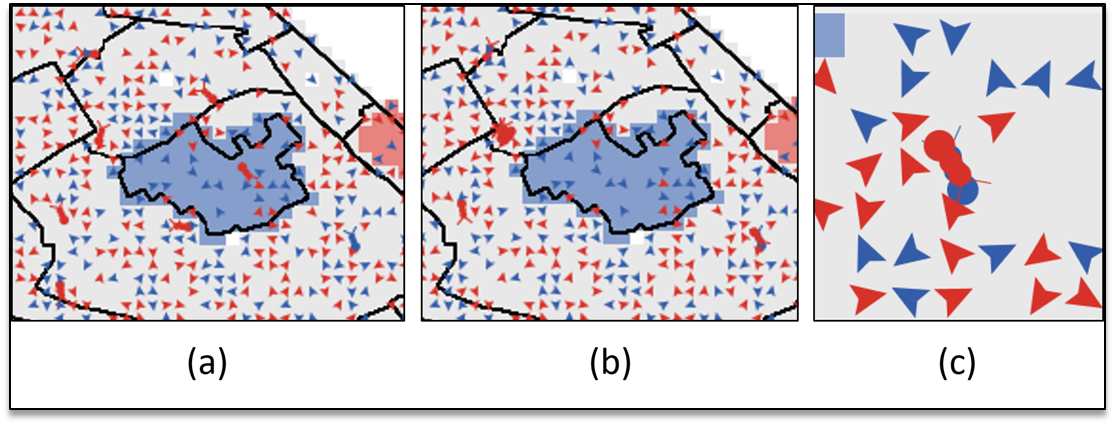}
\caption{Clustering of leaders. (a) A section of map with leaders (ant shape agents) placed randomly. (b) leaders cluster together creating nest (three of them are visible). (c) A closer view of one of the nest (the one on the right in (b)), with one cooperative and one fierce leader clustered.}
\label{fig:cluster}
\end{figure*}

\noindent During clustering, all leaders would choose a single cell in a specified range to cluster. Each leader on its turn chooses a cell in {\it cluster\_radius}, which has the maximum number of leaders placed on it and moves into that cell. If there are more than one options available, the leader can choose a cell randomly. If there is no such cell, the leader is placed in isolation; there would be no clustering. Clusters act as nests, ready for consolidation and inducing influence on agent population. A scenario describing clustering is exhibited in Fig. \ref {fig:cluster}.\\

\noindent {\bf Individual Fight}

In a nest, the fight between the leaders to claim the nest would initiate after reproduction and clustering. A fierce leader would initiate a fight if {\it prob\_init\_fight} is greater than a random floating point number. For example, if the value of {\it prob\_init\_fight} is $0.5$, there is an equal probability of fierce leader to initiate a fight or otherwise. If the fierce leader is set to initiate a fight, it will fight if there are more leaders in its nest. If the randomly chosen leader by the initiator is of type fierce as well, the probability for both of them winning the fight is equal. The leader losing the fight will die. If a fierce leader is imposing a fight on the cooperative leader, the probability of fierce leader dying is $60\%$, whereas, the probability of cooperative leader dying is $40\%$. All leaders in a nest perform the mechanism stated above until all other leaders have expired their turn or the leaders remaining are only cooperative, or there is no other leader (all died as a result of the fight). \\

\noindent {\bf Group Competition}

The fights in the nests would follow group competition. If two nests are close enough (identified by {\it radius\_competition}), the survival function {\it surv\_function} from Bartz and Hoelldobler \cite{bartz1982colony} would determine survival chances of the nest, which corresponds to the number of leaders living together in a nest. As a result of this competition, one of the nests (having chances of survival less than the other) would be destroyed along with the leaders in it. The cell representing a nest would first be assigned a value for the survival function {\it surv\_function}, calculated as \cite{bartz1982colony}:

\begin{equation} \label {eq:NFC}
{\it surv\_function} = -2.88 + 4.28 \times {\it nest\_pop} - 0.377 \times {\it nest\_pop}^{2}
\end{equation}
\noindent where {\it nest\_pop} is the number of leaders living together in a nest. Out of two competing nest, a nest having greater value of {\it surv\_function} would survive.\\

\noindent {\bf Influence}

The last step how a nest of leaders would influence the population of agents. All the nests would generate an infectious pheromone in an infection range around them, which is the product of {\it nest\_pop} and {\it radius\_competition}. The infection type is one of the randomly chosen leaders in the nest. A cell under this influence range would update its {\it infected\_with} value with the type of influence that this nest has: ``cooperation'', ``non-cooperation'', or ``null'' (if not in infection range of any of the nest). 

\subsection {Integrated Model}

Altogether, the conditions describing the state of an agent (being ``happy'' or ``unhappy'') in the context of segregation, not only depend on the population type in the neighborhood, but also, the aura of influence induced into the cellular space around it. The index of interest, the Desegregation\_Index is calculated as follows. 

Each agent in the population calculates its own {\it individual index of desegregation (IID)}, which is the ratio of the number of individuals NOT of its type, and the total number of neighbors in its neighborhood (8 adjacent cells of Moore's neighborhood \cite{kretz2007moore}). For example, if an individual is of type ``expat'' and if $3$ out of $5$ of its neighbors are also expats, the value of IID would be equal to $0.4$ ($40\%$ desegregation). Alternatively, if $2$ out of $5$ of its neighbors are also expats, the value of IID would be equal to $0.6$ ($60\%$ desegregation). Such an agent will be happy in the second case given that PDTU is $50\%$. 

IID is combined with nest influence to evaluate the state of happiness of an agent X. That is:

\begin{equation}
X = \left\{ \,
\begin{IEEEeqnarraybox}[][c]{l?s}
\IEEEstrut
``happy'' & if cond1 is true \\
``unhappy'' & if cond2 is true 
\IEEEstrut
\end{IEEEeqnarraybox}
\right.
\label{eq:example_left_right1}
\end{equation}  

where {\it cond1} is true when $IID >= {\it PDTU}$ AND ${\it infected\_with}_{c} =$ ``cooperation'' OR ``null''. This states the combined condition in which X's IID is high enough to suppress unhappiness and also, the nest influence supports desegregation (``cooperation'') or there is no influence (``null'' for no influencing nest in the surrounding); {\it cond2} is true otherwise. 

\section {Simulation and Result Analysis} \label {sec:sim}

\begin{table}[]
\caption{Desegregation\_Index}
\label{Table1}
\begin{tabular}{|l|l|l|l|l|l|l|l|}
\hline
\multicolumn{2}{|l|}{} & \multicolumn{3}{c|}{\textbf{NOL=25}} & \multicolumn{3}{c|}{\textbf{NOL=50}} \\ \hline
IR & FC & \multicolumn{1}{c|}{\textbf{PIF=0.1}} & \multicolumn{1}{c|}{\textbf{PIF=0.2}} & \multicolumn{1}{c|}{\textbf{PIF=0.5}} & \multicolumn{1}{c|}{\textbf{PIF=0.1}} & \multicolumn{1}{c|}{\textbf{PIF=0.2}} & \multicolumn{1}{c|}{\textbf{PIF=0.5}} \\ \hline
5 & 0.1 & 0.1287 & 0.1240 & 0.1156 & 0.1917 & 0.1701 & 0.1594 \\ \cline{2-8} 
& 0.2 & 0.1316 & 0.1199 & 0.1061 & 0.1908 & 0.1726 & 0.1611 \\ \cline{2-8} 
& 0.5 & 0.1172 & 0.1129 & 0.1008 & 0.1679 & 0.1595 & 0.1172 \\ \hline
25 & 0.1 & 0.0441 & 0.0458 & 0.0424 & 0.0631 & 0.0669 & 0.0536 \\ \cline{2-8} 
& 0.2 & 0.0441 & 0.0409 & 0.0369 & 0.0678 & 0.0569 & 0.0453 \\ \cline{2-8} 
& 0.5 & 0.0385 & 0.0385 & 0.0318 & 0.0531 & 0.0521 & 0.0469 \\ \hline
50 & 0.1 & 0.0432 & 0.0451 & 0.0394 & 0.0633 & 0.0521 & 0.0469 \\ \cline{2-8} 
& 0.2 & 0.0405 & 0.0353 & 0.0390 & 0.0636 & 0.0550 & 0.0532 \\ \cline{2-8} 
& 0.5 & 0.0402 & 0.0366 & 0.0349 & 0.0509 & 0.0509 & 0.0472 \\ \hline
\end{tabular}
\\
\newline NOL = Number of Leaders
\newline PIF = Probability of Initiating Fight
\newline IR = Leader Influence Rate
\newline FC = Fraction of Cooperation in the population.
\end{table}

\begin{table}[]
\caption{Happiness\_Index}
\label{Table2}
\begin{tabular}{|l|l|l|l|l|l|l|l|}
\hline
\multicolumn{2}{|l|}{} & \multicolumn{3}{c|}{\textbf{NOL=25}} & \multicolumn{3}{c|}{\textbf{NOL=50}} \\ \hline
IR & FC & \multicolumn{1}{c|}{\textbf{PIF=0.1}} & \multicolumn{1}{c|}{\textbf{PIF=0.2}} & \multicolumn{1}{c|}{\textbf{PIF=0.5}} & \multicolumn{1}{c|}{\textbf{PIF=0.1}} & \multicolumn{1}{c|}{\textbf{PIF=0.2}} & \multicolumn{1}{c|}{\textbf{PIF=0.5}} \\ \hline
5 & 0.1 & 0.6663 & 0.6774 & 0.7018 & 0.5006 & 0.5585 & 0.5869 \\ \cline{2-8} 
& 0.2 & 0.6936 & 0.6882 & 0.7245 & 0.5033 & 0.5553 & 0.5841 \\ \cline{2-8} 
& 0.5 & 0.6936 & 0.7065 & 0.7363 & 0.5638 & 0.5879 & 0.6901 \\ \hline
25 & 0.1 & 0.8925 & 0.8887 & 0.8958 & 0.8512 & 0.8414 & 0.8747 \\ \cline{2-8} 
& 0.2 & 0.8916 & 0.9003 & 0.9110 & 0.8423 & 0.8683 & 0.8956 \\ \cline{2-8} 
& 0.5 & 0.9063 & 0.9063 & 0.9216 & 0.8768 & 0.8808 & 0.8921 \\ \hline
50 & 0.1 & 0.8937 & 0.8891 & 0.9021 & 0.8520 & 0.8808 & 0.8921 \\ \cline{2-8} 
& 0.2 & 0.9017 & 0.9148 & 0.9058 & 0.8519 & 0.8698 & 0.8742 \\ \cline{2-8} 
& 0.5 & 0.9026 & 0.9112 & 0.9153 & 0.8823 & 0.8820 & 0.8894 \\ \hline
\end{tabular}
\\
\newline NOL = Number of Leaders
\newline PIF = Probability of Initiating Fight
\newline IR = Leader Influence Rate
\newline FC = Fraction of Cooperation in the population.
\end{table}

\begin{figure*} 
\centering
\includegraphics[width=0.99\textwidth]{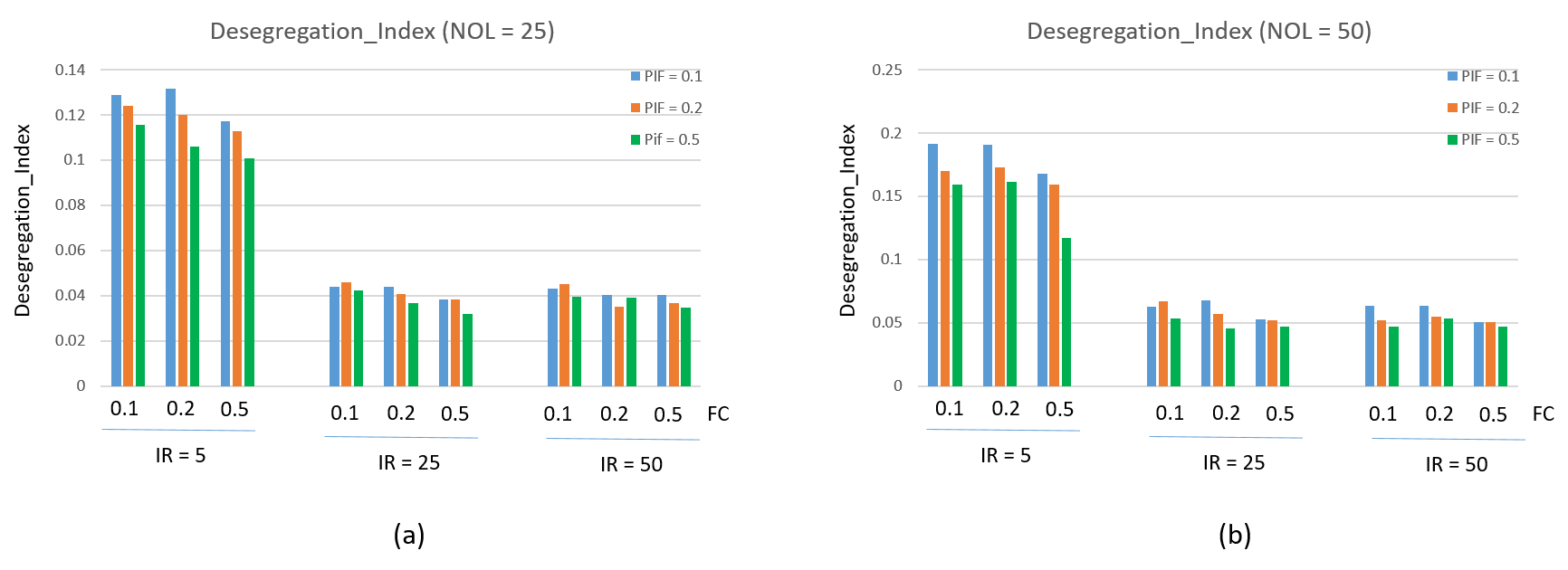}
\caption{Average Desegregation\_Index of Simulation runs corresponding to TABLE I. NOL = Number of Leaders, PIF = Probability of Initiating Fight, IR = Leader Influence Rate, FC = Fraction of Cooperation in the population.}
\label{fig:desegindex}
\end{figure*}

\begin{figure*} 
\centering
\includegraphics[width=0.99\textwidth]{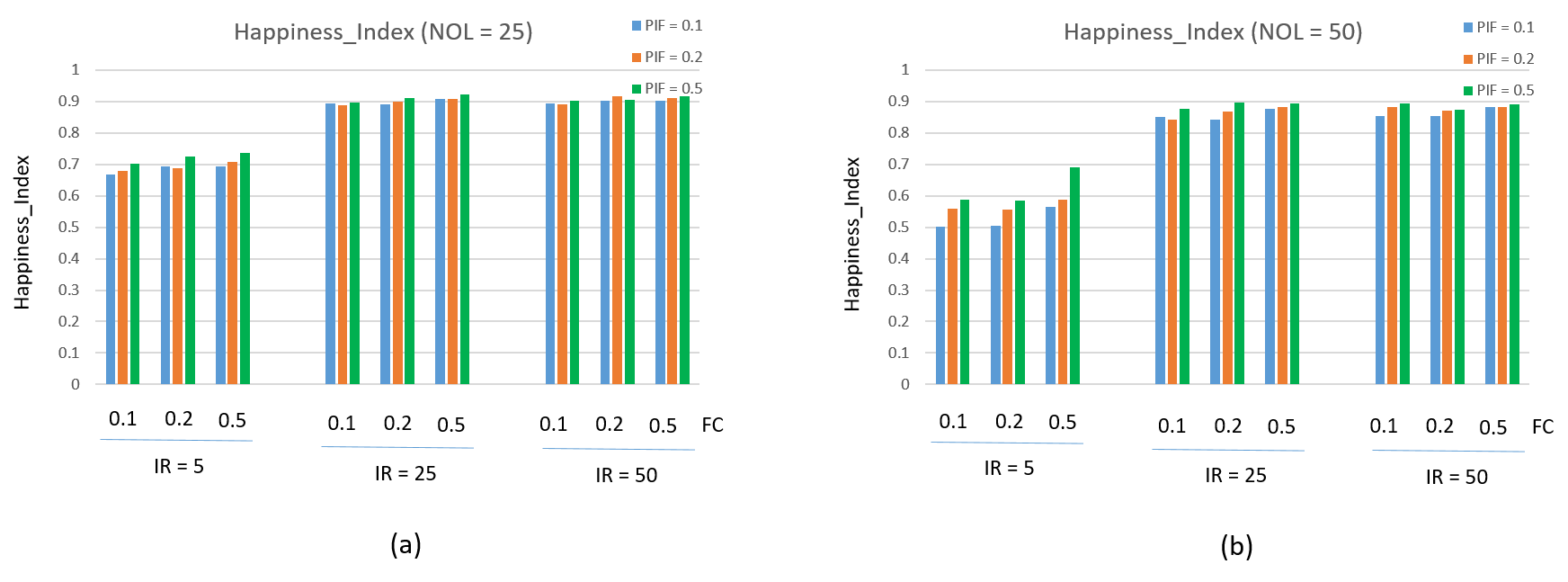}
\caption{Average Happiness\_Index of Simulation runs corresponding to TABLE II. NOL = Number of Leaders, PIF = Probability of Initiating Fight, IR = Leader Influence Rate, FC = Fraction of Cooperation in the population.}
\label{fig:happyindex}
\end{figure*}

\begin{figure*} [!t] 
\centering
\includegraphics[width=0.99\textwidth]{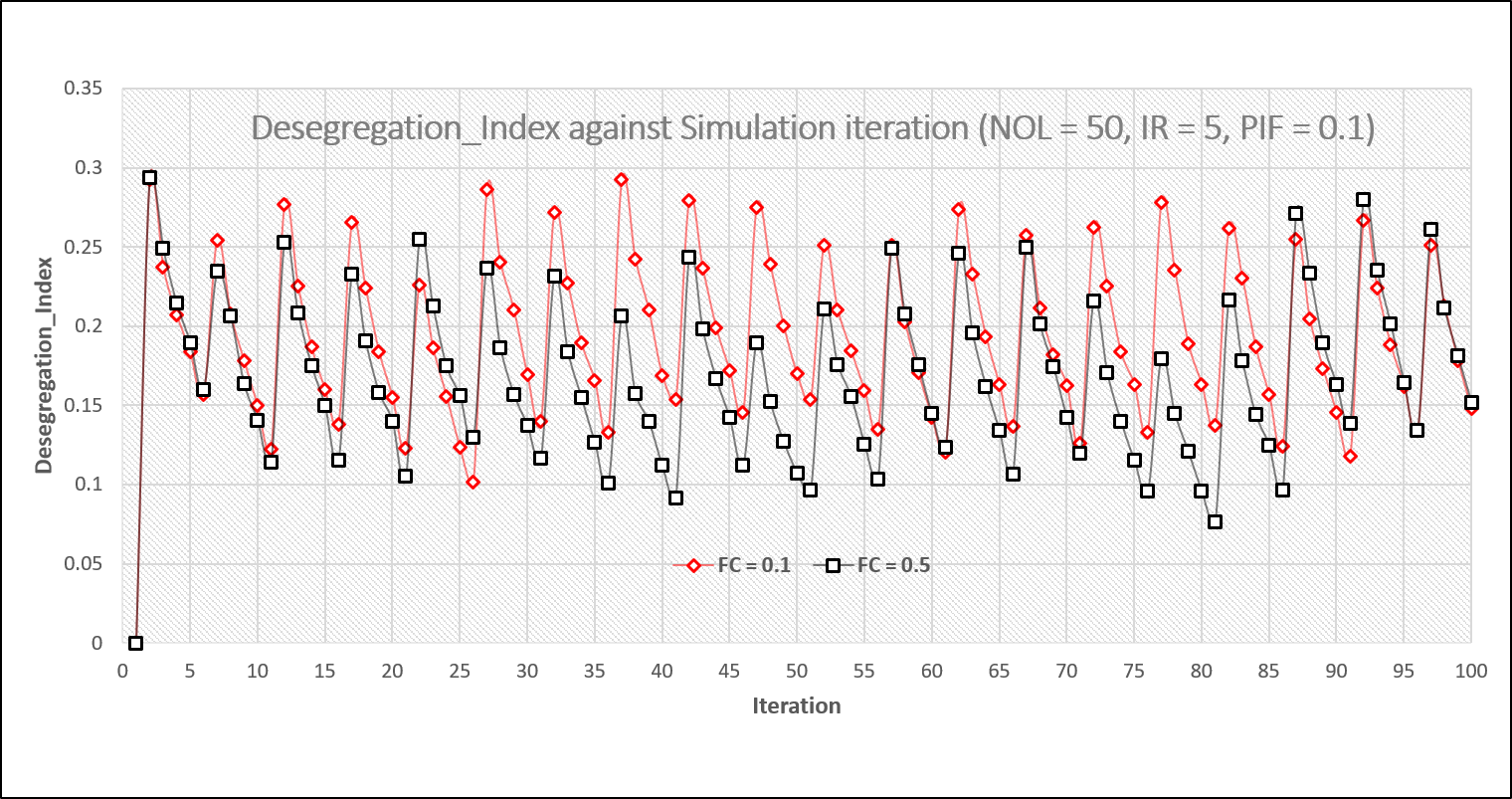}
\caption{Comparison of desegregation index against simulation iterations between FC = 0.1 and FC = 0.5. Specific case of NOL = 50, IR = 5 and PIF = 0.1. [NOL = Number of Leaders, PIF = Probability of Initiating Fight, IR = Leader Influence Rate, FC = Fraction of Cooperation in the population]}
\label{fig:g1}
\end{figure*}

For simulation, Netlogo \cite{tisue2004netlogo}, a CA-based agent-based modeling tool is used. The simulation is executed for a real city (Sohar, Oman) using its GIS map. Several important cases across multiple settings of simulation parameters are analyzed. Parameters used in the simulation are as follows:

\begin{itemize}
\item {\bf NOL:} number of leaders constituting the virtual layer. The values are 25 and 50.
\item {\bf FC:} fraction of leaders of type cooperative in the entire population of leaders (NOL). The values are 0.1, 0.2, and 0.5.
\item {\bf IR:} leaders' influence rate, which represents how frequently leaders are influencing the population concerning simulation time lag. Values are every 5th, 25th and 50th iteration.
\item {\bf PIF:} probability of initiating the fight by a fierce leader, representing the intensity of influence of group leaders. Values are 10\%, 20\%, and 50\%.
\end{itemize} 

The other parameters related to the model of influence are constants:
\begin{itemize}
\item pmutation: mutation probability of each generation (set to 1\%).
\item cluster-radius: radius of the cluster (set to 10 cells).
\item radius-competition: the radius of the group competition (set to 50 cells).
\end{itemize}

The constant parameters related to the model of segregation (desegregation) are:
\begin{itemize}
\item population size: set to 5000 agents.
\item segregation threshold: set to 40\%.
\item PDTU: difference in population types in the neighborhood of an agent (the -ve percentage that makes it unhappy). set to 40\%.
\end{itemize}

It is worth noting that scenarios representing different values for constants of ranges, population, and thresholds would entirely change the simulation results. Therefore, these values are kept constant, and only those, which give a consistent meaning to phenomena of interest are varied.

Table \ref{Table1} and Table \ref{Table2} outline the simulation results. To calculate the global Desegregation\_Index, IID of all the agents is accumulated and divided by the total number of agents. Similarly, Happiness\_Index, which is inversely proportional to IID, is an accumulated global value of agent happiness as formulated in the equation \ref{eq:example_left_right1}. Consequently, with an increase in the desegregation index, the population as a whole is expected to be less happy. This is exemplified by the value of the index in case of the model of segregation applied alone. Without virtual leaders, and their influence, the index is equal to $0.0152$. In fact, this represents that the population is segregated, and index of desegregation is $0.0152$. It understandably makes overall population very happy with an index equal to $0.9568$.
\begin{figure*} 
\centering
\includegraphics[width=0.75\textwidth]{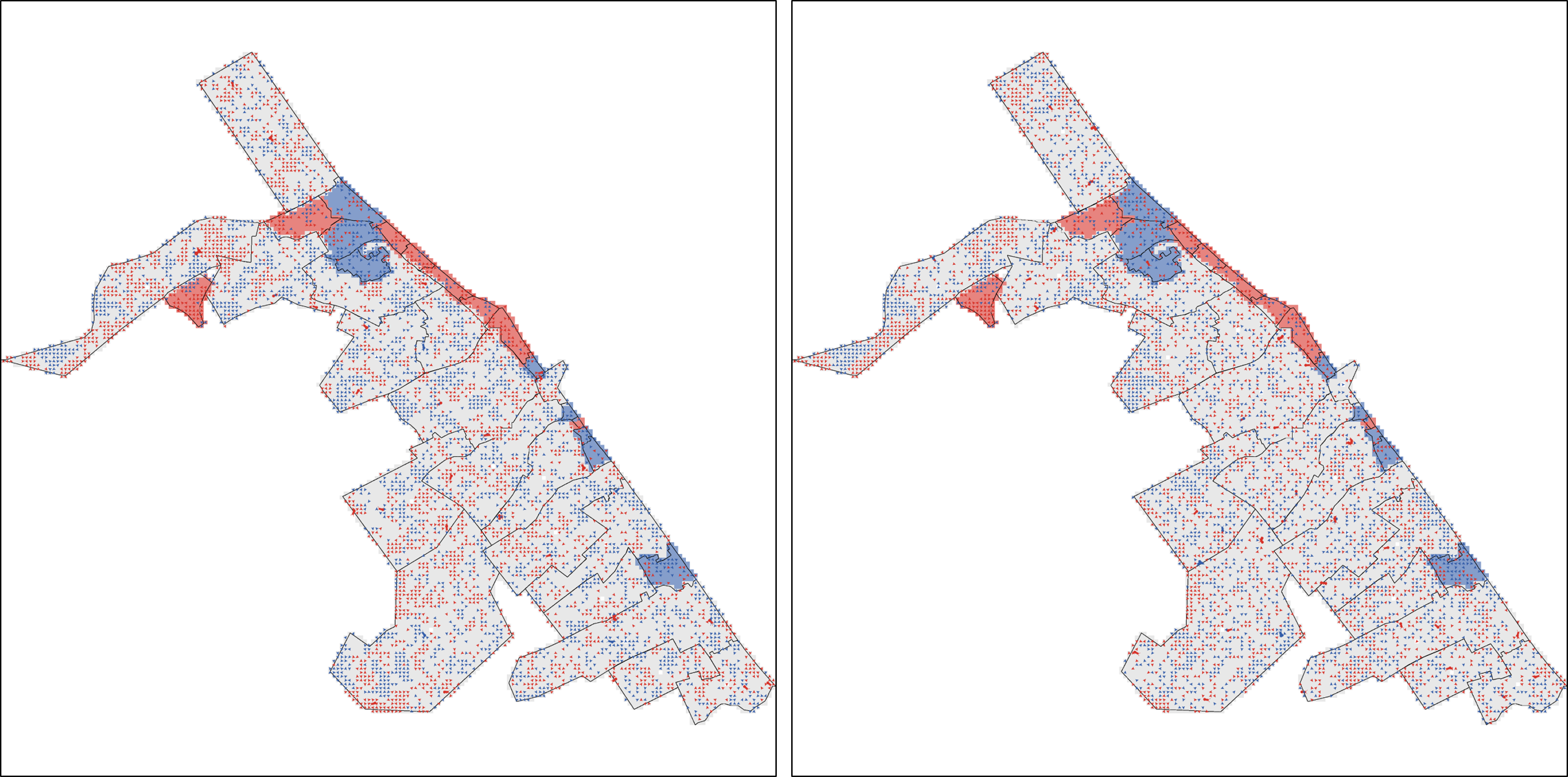}
\caption{Effect of Leaders' influence on the population represented by simulation screenshot at iteration 30 (left) and 31 (right) in case NOL = 50, IR = 5, PIF = 0.1, FC = 0.1. [NOL = Number of Leaders, PIF = Probability of Initiating Fight, IR = Leader Influence Rate, FC = Fraction of Cooperation in the population]}
\label{fig:ss1}
\end{figure*}

Analyzing Table \ref{Table1}, and the corresponding graphs in Fig. \ref{fig:desegindex}, it is evident that with an increase in the number of leaders, the Desegregation\_Index increases, that is, the population gets desegregated. It is seen that with more frequency of influence dissemination, the population gets desegregated as well. Overall, with an increase in the intensity of the influence, the value of desegregation decreases. It is because the fierce leaders kill cooperative leaders, thus decreasing the corresponding influence on the population. The maximum desegregation is achieved when the intensity of the influence as well as the fraction of cooperative leaders are minimal. Happiness\_Index corresponds to Desegregation\_Index, and all values are consistent as shown in Table \ref{Table2}, and corresponding graphs in Fig. \ref{fig:happyindex}.

\begin{figure*} 
\centering
\includegraphics[width=0.75\textwidth]{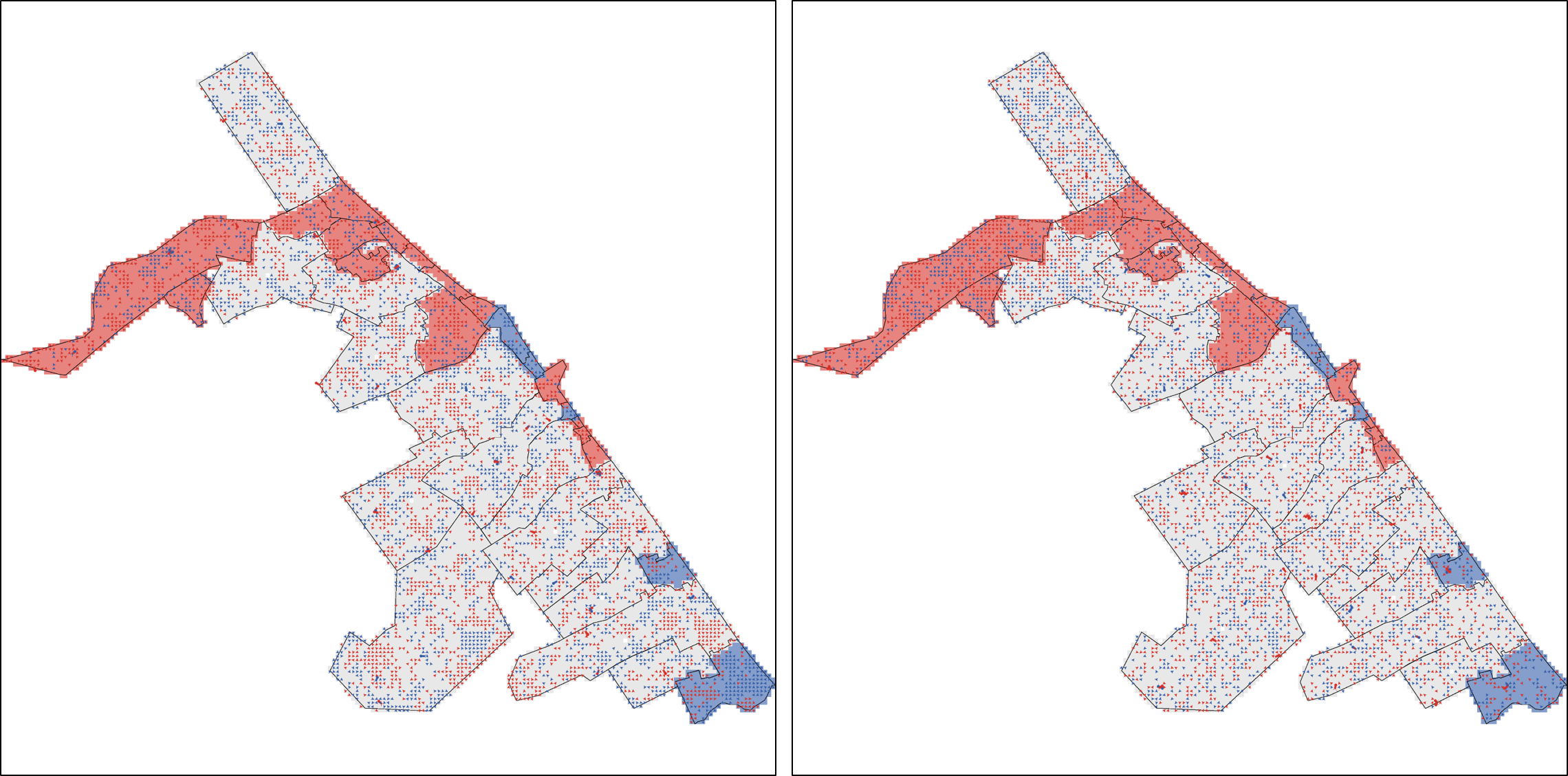}
\caption{Effect of Leaders' influence on the population represented by simulation screenshot at iteration 30 (left) and 31 (right) in case NOL = 50, IR = 5, PIF = 0.1, FC = 0.5. [NOL = Number of Leaders, PIF = Probability of Initiating Fight, IR = Leader Influence Rate, FC = Fraction of Cooperation in the population]}
\label{fig:ss2}
\end{figure*}

\begin{figure*} 
\centering
\includegraphics[width=0.99\textwidth]{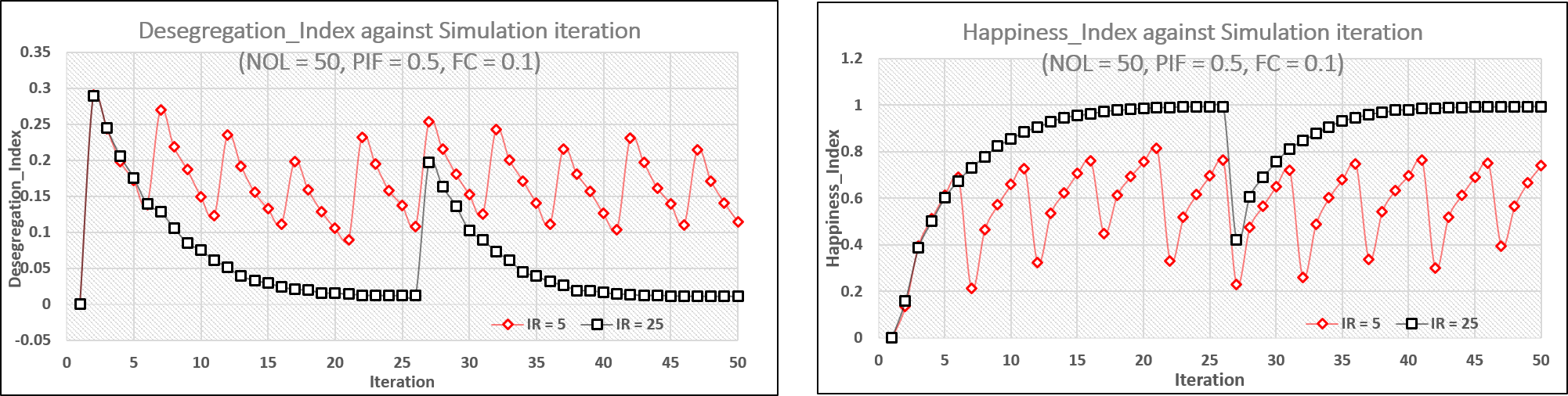}
\caption{Comparison of desegregation index (left) and happiness index (right) between IR = 5 and IR = 25 in case NOL = 50, FC = 0.1, PIF = 0.5. [NOL = Number of Leaders, PIF = Probability of Initiating Fight, IR = Leader Influence Rate, FC = Fraction of Cooperation in the population]}
\label{fig:g2}
\end{figure*}

\begin{figure*} 
\centering
\includegraphics[width=0.75\textwidth]{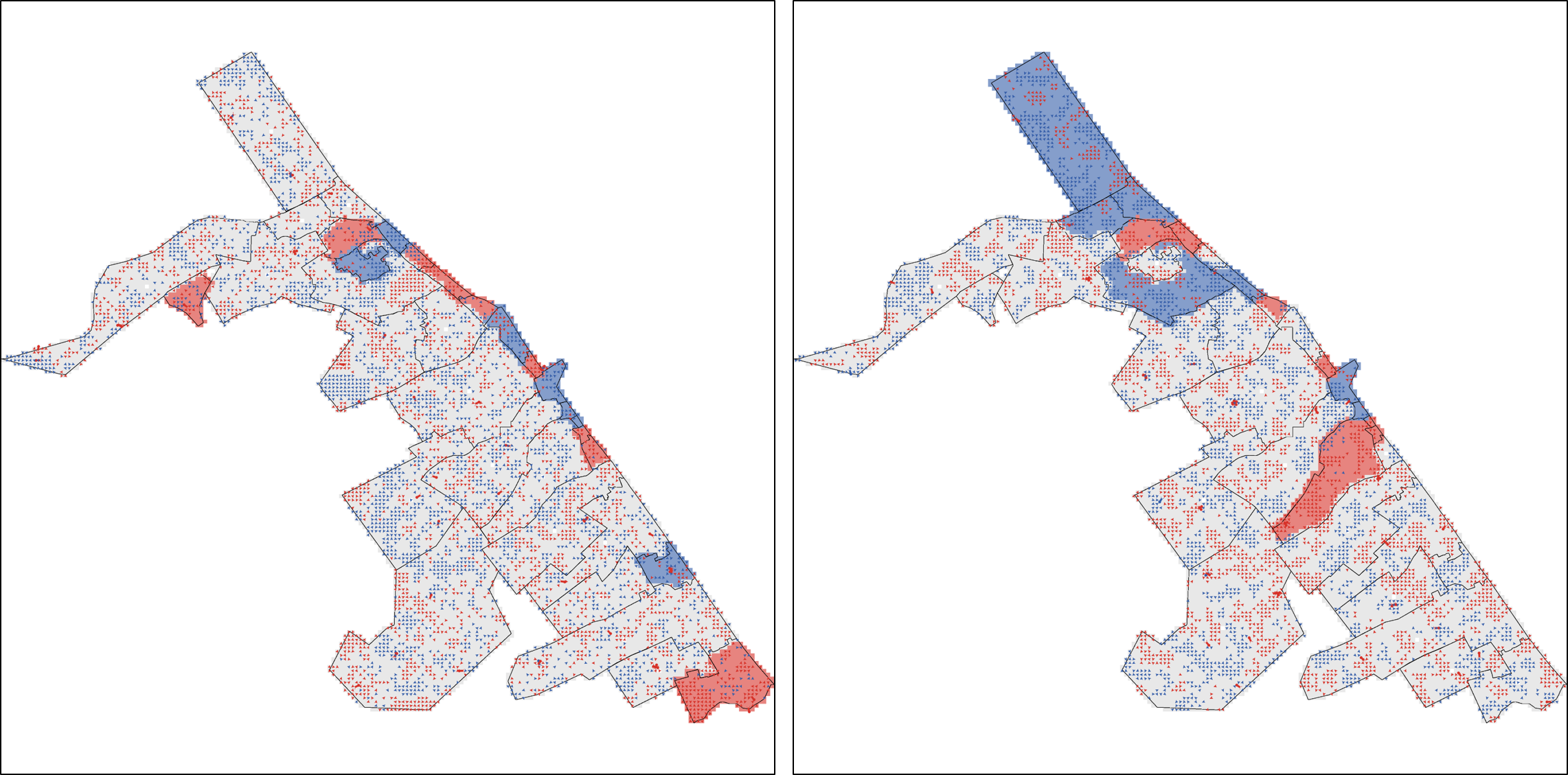}
\caption{Simulation screenshot at iteration 25, just before invocation of periodic influence cycle for IR = 5 (left) and IR = 25 (right) representing the importance of more periodic group influence for population desegregation, in case NOL = 50, PIF = 0.1, and FC = 0.1. [NOL = Number of Leaders, PIF = Probability of Initiating Fight, IR = Leader Influence Rate, FC = Fraction of Cooperation in the population]}
\label{fig:ss3}
\end{figure*}

The maximum desegregation equal to $0.1917$ is achieved when $NOL = 50, IR = 5, PIF = 0.1,$ and $FC = 0.1$. However, there is a difference of almost $15\%$, when $FC = 0.1$ with $FC = 0.5$ is compared. A closer look at this trend exhibited in the graph in Fig. \ref{fig:g1}. The graph shows a regular pattern. Overall, the level of desegregation is highest when a new phase of influence dissemination is started (after each five simulation iterations, in this case, i.e. IR = 5), and decreases continuously afterwards unless it reaches a new peak at the wake of a new influence dissemination phase. In each such phase, the index is higher when FC is less, which means more desegregation. It means that more cooperative leaders act negatively for the desegregation. It is an interesting result, which is justified using simulation screenshots given in Fig. \ref{fig:ss1} and Fig. \ref{fig:ss2}.

Fig. \ref{fig:ss1} (left) shows the arrangement of the agents at iteration 30, one iteration before the leaders assert their influence. It is evident that there are few cooperative leaders. Consequently, the population is quite segregated (except for some isolated islands where cooperative leaders can assert their influence). At iteration 31 (right), the population gets quite desegregated due to the refreshed influence of the leaders. However, it does not happen at the fringes of the city due to the absence of cooperative leaders. These screenshots represent situations when $FC = 0.1$. A similar analysis is done when $FC = 0.5$ as shown in Fig. \ref{fig:ss2}. There is not much difference in arrangement of the agents at iteration 40 (left), one iteration before the leaders assert their influence. However, at iteration 41, the reach of leaders at the fringes of the city (due to more number) is evident.
Hence, this strange outcome relates to demographic features of the population, not the comparative number of leaders, and their influence.       

Another interesting result is about comparing the desegregation and happiness in case of varying influence periodicity. In the constant conditions of $NOL = 50, PIF = 0.5, FC = 0.1$, more periodic influence dissemination keeps the desegregation at a higher level. It is shown in the graph in Fig. \ref {fig:g2} (left). At the initial stages of the simulation, both for $IR = 5$ and $IR = 25$, the Desegregation\_Index is similar (around 0.3), which drops for subsequent iterations following the similar trend. However, when the first period of refreshing of the influence happens (at iteration 5) for $IR = 5$, the Desegregation\_Index elevates from around 0.15, while for $IR = 25$, the index keeps dropping. In the subsequent periods, each separated by five timestamps, the Desegregation\_Index keeps fluctuating between a maximum of around 0.26 and a minimum of around 0.1. In case of $IR = 25$, the Desegregation\_Index almost drops to 0.1 and then gets a boost at iteration 26 when the first period of refreshing of the influence happens for $IR = 25$. After refresh, the peak value does not surpass 0.2. However, the real issue is the length of time for which the value of index remains between 0.1 and 0.0, which is almost 80\% of the time. Opposite of Desegregation\_Index is the pattern of Happiness\_Index shown in the graph in Fig. \ref {fig:g2} (right).  

The above pattern is evident from the simulation screenshots given in Fig. \ref{fig:ss3}, evidencing the importance of increased periodicity of influence for the population desegregation.

Overall, the simulation results reveal the following outcomes:

\begin {itemize}
\item With the increase in the frequency of contact of the leaders with the population, desegregation increases.
\item With the increase in the intensity of contact of the leaders with the population, desegregation decreases. 
\item Desegregation is only possible if the population gets in contact with the leaders.
\end {itemize}

In the next section, a discussion on these outcomes is presented.

\section {Discussion and Conclusion} \label {sec:conc}

Contradictory views regarding the effect of contact on desegregation are available. On one extreme, the contact hypothesis \cite{miller2013groups} states that: ``the idea that prejudice and hostility between members of segregated groups can be reduced by promoting the frequency and intensity of intergroup contact''. Whereas, one the other extreme, it is stated that \cite{denis2015contact} ``contact tends to reproduce, rather than challenge, the inequitable racial structure''. However, both hypotheses do not take into account the contradictions and sustainability of influencing groups. Since, segregation is a result of internal bias of individuals, so must be the counter of it, i.e. desegregation. Social bias is not purely an internal factor. It consolidates with the passage of time, gained with experience and social influence. Hence, the aversion of bias induced by some imposing group should not be examined as a deterministic uni-dimensional factor acting against segregation.

In this paper, a model based on group selection \cite{shaffer2016foundress} is customized to integrate it with the model of group influence on the population. The model resolves the Foundress Dilemma for each generation of spatial group leaders and provides an evolutionary mechanism that may transform the virtual population of influence. The real population, in addition of having internal bias, gets influenced by an external non-deterministic bias, generated by virtual group leaders.   

The emerging nature of group influence integrated with population decision (to desegregate or not to) produces more knowledgeable outcomes, validating some aspects of the theories and hypothesis outlined above, while rejecting others. It can be argued that contact itself (without any external anti-segregation impact) is not enough to change racial structure (as suggested in \cite{denis2015contact}). However, nobody would have any such expectation, as this is against pure human nature. With external influence, the change in racial structure is possible, thus, providing a justification of keeping real population and leaders as two separate groups (layers).

A part of contact hypothesis is validated, i.e. {\it with the increase in the frequency of contact of leaders with the population, desegregation increases}. However, with a strong assumption that the virtual population is analogous to the real population (predominately favoring segregation), the second part of the hypothesis is not validated. In fact, it was observed that {\it with the increase in the intensity of contact of leaders with the population, desegregation decreases}. More intense the viewpoint of virtual leaders (influence leaders) is, more intense would be its impact, practically, favoring segregation instead of desegregation.  

The simulation results reveal that it is also not true that decrease in the intensity of contact always increase the desegregation. A decrease in the intensity of contact combined with decrease in the number of cooperative leaders favoring desegregation results in best case with maximum desegregation. 

\noindent {\bf Future Work:} Although the simulation results are validated against two contradictory theories of contact, validating them on real city population will be an interesting dimension to consider for the future work.

\ifCLASSOPTIONcaptionsoff
  \newpage
\fi



%

\bibliographystyle{ieeetran}
\bibliography{main}

%
%

%

\begin{IEEEbiography}[]{Kashif Zia}
Kashif Zia is currently working as Assistant Professor at Sohar University, Oman. He obtained his PhD from the Institute for Pervasive Computing, Johannes Kelper University in Linz, Austria. Zia’s research interests revolve around socio-technical systems and computational social science with particular focus on crowds dynamics and smart cities. His PhD work was related to large-scale agent-based modeling and analysis of socio-technical systems utilizing parallel and distributed simulation.
\end{IEEEbiography}

\begin{IEEEbiography}[]{Dinesh Kumar Saini}
Dr. Dinesh Kumar Saini received a PhD degree in 2005. Dr. Saini hold the rank of an Associate Professor at Sohar University, Oman and an Adjunct Associate and Research Fellow with the University of Queensland, Brisbane Australia. Research interest in Software Engineering, Cyber Defense and Mathematical Modeling, Content Management Systems, and Computational Social Science.
\end{IEEEbiography}


\begin{IEEEbiography}[]{Arshad Muhammad}
Arshad Muhammad received a PhD degree in 2010 Computer Science from Liverpool John Moores University, UK. From 2010 till 2013, worked as Assistant Professor with City University, Pakistan. In 2013, he joined Sohar University, Oman as Assistant Professor. He has more than 8 years of experience in research and academics. His areas of expertise are Peer-to-Peer networks, Networked Appliances, Agent-based modeling \& simulation, IoT, and Service-Oriented Architectures.
\end{IEEEbiography}

\begin{IEEEbiography}[]{Alois Ferscha}
Alois Ferscha received a Mag. degree in 1984, and a PhD in business informatics in 1990, both from the University of Vienna, Austria. From 1986 through 2000 he was with the Department of Applied Computer Science at the University of Vienna at the levels of Assistant and Associate Professor. In 2000 he joined the University of Linz as full Professor where he is now head of the department for Pervasive Computing and the speaker of the JKU Pervasive Computing Initiative. Currently he is focused on pervasive and ubiquitous computing, networked embedded systems, embedded software systems, wireless communication, multiuser cooperation, distributed interaction and distributed interactive simulation.
\end{IEEEbiography}

\vfill


\end{document}